\documentclass{an}
\usepackage{graphicx}
\usepackage{times}
\Volume{00}              
\Year{0000}              
\Month{00}               
\Pagespan{000}{000}      

\begin{document}
\lhead[\thepage]{M.J. Page et al.: X-ray and optical properties of faint X-ray
sources}
\rhead[Astron. Nachr./AN~{\bf XXX} (200X) X]{\thepage}
\headnote{Astron. Nachr./AN {\bf 32X} (200X) X, XXX--XXX}

\title{X-ray and optical properties of X-ray
sources in the 13hr XMM-Newton/Chandra deep survey}

\author{M.J.~Page\inst{1} \and 
I.M.~M$\rm^{c}$Hardy\inst{2} \and K.F.~Gunn\inst{2} 
\and 
N.S.~Loaring\inst{1} \and K.O.~Mason\inst{1} \and T.~Sasseen\inst{3} \and 
A.~Newsam\inst{4} \and A.~Ware\inst{3} \and J.~Kennea\inst{3} \and
 K.~Sekiguchi\inst{5} \and T.~Takata\inst{5}}
\institute{Mullard Space Science Laboratory, University College London,
 Holmbury St. Mary, Dorking, Surrey, RH5 6NT, UK
\and Department of Physics and Astronomy, University of Soutampton, Southampton
 SO17, 1BJ, UK
\and Physics Department, University of California, Santa Barbara, CA, 93106,
 USA
\and Liverpool John Moores University, Astrophysics Research Institute,
 Birkenhead L41 1LD, UK
\and Subaru Telescope, National Astronomical Observatory of Japan,
 650 North A'ohoku Place, Hilo, HI 96720, USA}
\date{Received {date will be inserted by the editor}; 
accepted {date will be inserted by the editor}} 

\abstract{ The 13hr {\em XMM-Newton}/{\em Chandra} 
deep survey is the first of two
extremely deep {\em XMM-Newton} 
fields observed by the XMM-OM consortium. A 120 ks
{\em Chandra}
 mosaic, covering 0.2~deg$^{2}$, provides sensitive, confusion-free 
point source detection with
sub-arcsecond positions, while the 200 ks {\em XMM-Newton} 
observation provides high
quality X-ray spectroscopy over the same sky area. 
We have optical spectroscopic identifications for
70 X--ray sources. Of these, 42 are broad emission-line AGN with a wide 
range of
redshifts. The optical counterparts of a further 23 sources are
 narrow
emission line galaxies and absorption line galaxies. These 23 
sources all lie at
$z<1$ and 
typically have lower X--ray luminosities than the broad-line AGN. 
About half of them
show significant X--ray absorption and are almost certainly intrinsically
absorbed AGN. However some of them have unabsorbed, AGN-like, power-law
components in their X--ray spectra, but do not show broad emission lines in
their optical spectra. These sources may be weak, unobscured AGN in bright 
galaxies and
their existence at low redshifts could be a consequence of the strong
cosmological evolution of AGN characteristic luminosities.  
\keywords{surveys
--- X--rays --- galaxies: active --- galaxies: evolution} }
\correspondence{mjp@mssl.ucl.ac.uk}

\maketitle

\section{Introduction}

Above 1 keV, the cosmic X--ray background (CXRB) is predominantly the 
integrated
emission of many discrete sources. In the 0.5-2 keV X--ray band, the deepest
surveys with {\em ROSAT} resolved 70-80\% of the X--ray background into
individual sources, the majority of which are broad-line AGN (e.g.  Hasinger et
al. \nocite{hasinger98} 1998, McHardy et al. \nocite{mchardy98} 1998). 
However the energy density of the CXRB peaks at much
higher energy, $\sim 30$ keV, and to reproduce its broadband spectral shape, 
CXRB synthesis models require a large population of hard-spectrum X--ray
sources. The most enduring of these synthesis models invoke a large population
of absorbed AGN, as expected in AGN `unification schemes' (Setti and Woltjer
\nocite{setti89} 1989, Comastri et al. \nocite{comastri95} 1995, Gilli,
Risaliti \& Salvati \nocite{gilli99} 1999).
Deep surveys performed with {\em XMM-Newton} and {\em Chandra} have now 
succeeded
in resolving a large fraction of the CXRB up to 10 keV (e.g. Brandt et al.,
\nocite{brandt01} 2001, Campana et
al. \nocite{campana01} 2001, Hasinger et al., \nocite{hasinger01} 2001). 
The X-ray and
optical properties of the {\em Chandra} and {\em XMM-Newton} sources will allow us
to test these synthesis models.

The 13hr deep field is a project to investigate the astrophysics of the major
contributors to the CXRB, particularly sources around the break
in the X--ray source counts where the contribution to the CXRB per log-flux
interval peaks. In order to understand the phenomena, processes and conditions
in these sources we combine the high quality X--ray spectra of 
{\em XMM-Newton}
with the precise positions of {\em Chandra} and a host of other 
multi-wavelength data. The 13hr field was the location of one of the deepest
{\em ROSAT} surveys (McHardy et al. \nocite{mchardy98} 1998) and is
a region of extremely low Galactic absorption (${\rm N_{H}} \sim 6.5 \times
10^{19}$~cm$^{-2}$). 

\section{Observations}

The 13hr deep field was observed with four separate {\em Chandra} ACIS-I
pointings, resulting in a
mosaic of {\em Chandra} images which covers 0.2~deg$^{2}$. 
A total of 214 point sources are detected to a limiting flux of $1.3 \times
10^{-15}$~erg~cm$^{-2}$~s$^{-1}$ in the 0.5-7 keV
energy band, and for the majority of these positions are accurate to $<$ 
1$''$ (McHardy et al. \nocite{mchardy03} 2003).

{\em XMM-Newton} observations totalling 200 ks were performed in June 2001.
After processing with the {\em XMM-Newton} SAS, the EPIC event lists were
time-filtered to remove background flares resulting in 130ks of good data.
Images and exposure maps for the three EPIC cameras combined were constructed
in the energy bands 0.2-0.5 keV, 0.5-2.0 keV, 2.0-5.0 keV and 5.0-12.0
keV. Source searching was performed with the SAS {\small EBOXDETECT} and
{\small EMLDETECT} tasks. Background maps were constructed by excising the
sources from the images and then performing a maximum-likelihood fit with
uniform, vignetted and unvignetted backgrounds. Several iterations of
source-detection, source-excision and background fitting were performed to
optimise the source detection. The on-axis flux limits in the individual 
energy bands are $5\times 10^{-16}$ erg~cm$^{-2}$~s$^{-1}$ (0.2-0.5 keV),
$5\times 10^{-16}$ erg~cm$^{-2}$~s$^{-1}$ (0.5-2.0 keV),
$1.4\times 10^{-15}$ erg~cm$^{-2}$~s$^{-1}$ (2-5 keV), and
$1\times 10^{-14}$ erg~cm$^{-2}$~s$^{-1}$ (5-12 keV).
In total 216 sources are detected in the {\em XMM-Newton} images; 
three quarters of the
sources are detected independently with both {\em XMM-Newton} 
and {\em Chandra}. 
123 {\em XMM-Newton} 
sources have $> 100$ counts, and therefore have useful X-ray spectra.

\begin{figure}
\resizebox{\hsize}{!}
{\includegraphics[angle=270]{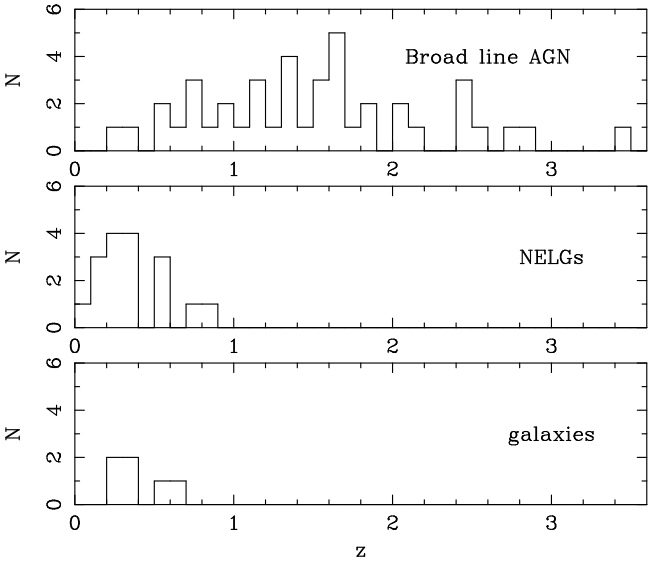}}
\caption{Redshift distribution for the extragalactic sources.}
\label{fig:redshifts}
\end{figure}

\begin{figure}
\resizebox{\hsize}{!}
{\includegraphics[width=6.5cm]{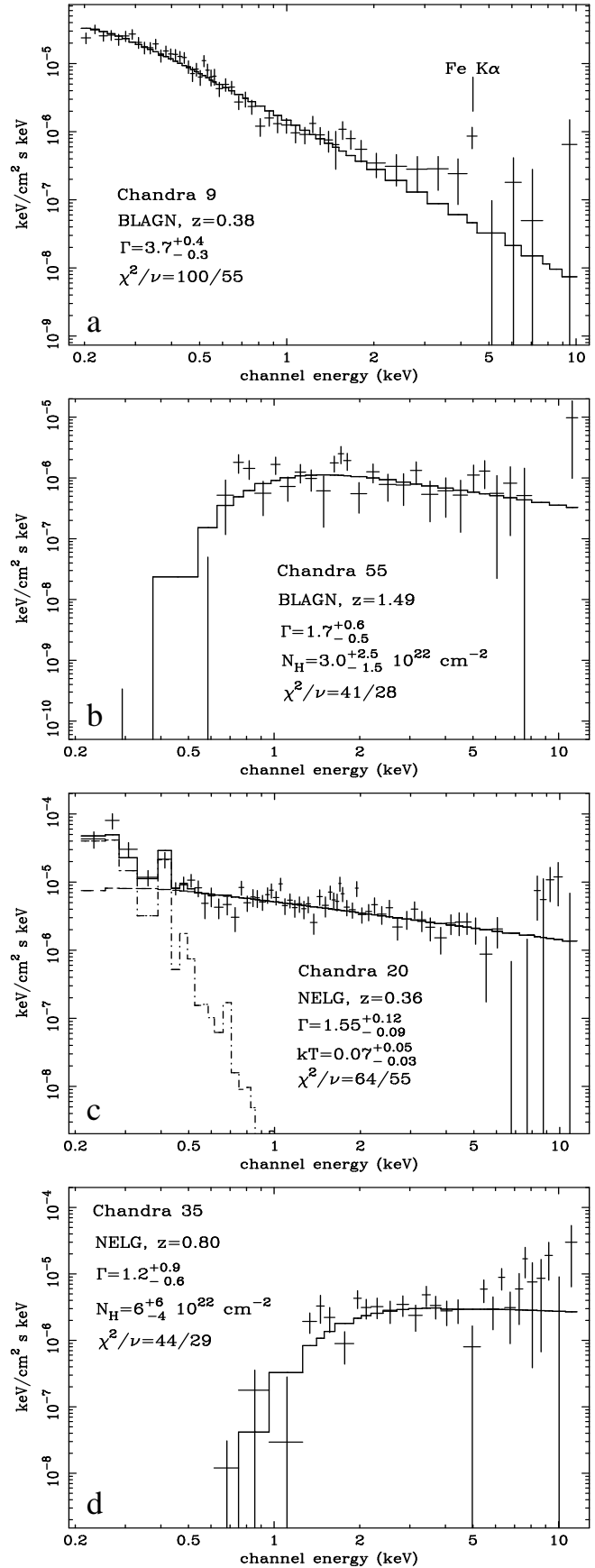}}
\caption{Some example X--ray spectra from the 13hr field. From the top, a) an
ultra-soft BLAGN, b) a BLAGN with a heavily absorbed X--ray spectrum, c) a NELG
with a multi-component X-ray spectrum, d) a NELG which harbours a powerful but
heavily absorbed AGN.}
\label{fig:spectra}
\end{figure}

In order to identify the optical counterparts to the X--ray sources we have 
taken deep
images on the Subaru Telescope, the William Herschel Telescope, the Isaac
Newton Telescope, and the 4m at
Kitt Peak. The deepest imaging comes from Suprime-Cam on Subaru and
reaches $R\sim$27. About 80\% of the X-ray sources have optical counterparts
with $R<24$ and with few exceptions these counterparts are unambiguous. 
We recently
obtained multi-object optical spectroscopy with AF2/WYFFOS on the WHT and with
LRIS on Keck. Combined with the existing optical spectra which were taken as 
part of the
{\em ROSAT} survey follow up, we have optical identifications and redshifts for
70 X-ray sources, most of which have $R<23$.

\begin{figure*}
\resizebox{\hsize}{!}
{\includegraphics[angle=270]{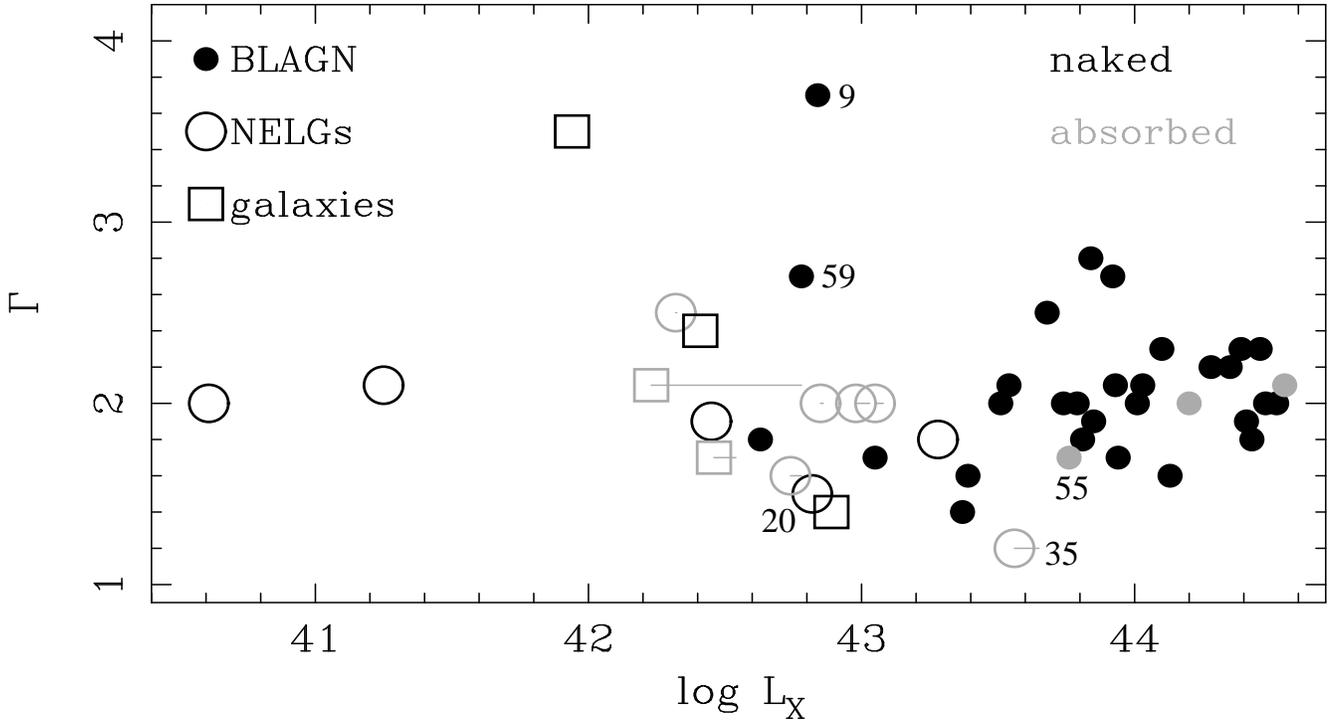}}
\caption{Best fit spectral slopes and 2-5 keV luminosities for the optically 
identified
sources with good {\em XMM-Newton} spectra. Absorbed sources are shown in grey,
and the horizontal lines show where the sources lie after their 
luminosities are corrected for absorption. The sources shown in
Figs. \ref{fig:spectra} and \ref{fig:chandra59} are marked with their {\em
Chandra} numbers.}
\label{fig:specresults}
\end{figure*}

\section{Optical and X-ray spectroscopic properties}
\label{sec:properties}

\begin{table}[h]
\caption{Optical spectroscopic properties of the {\em Chandra} sources. 
The ratio of
absorbed:unabsorbed sources is based on fits to the {\em XMM-Newton} spectra
with the absorbed source
criterion given in Section \ref{sec:properties}.}
\label{tab:IDs}
\begin{tabular}{lcccc}\hline
Optical & Number & $\langle z \rangle$& $\langle \Gamma \rangle$ & ratio \\ 
classification & &                    & & absorbed:unabsorbed \\
\hline
BLAGN & 42 & 1.52 & 2.1 & 3:29\\
NELGs & 17 & 0.36 & 1.9 & 6:5\\
Galaxies & 6 & 0.41 & 2.2 & 2:3\\
Stars & 5 & - & - & - \\
\hline
\end{tabular}
\end{table}

We have divided the extragalactic spectroscopically-identified {\em Chandra} 
sources
according to their emission line properties. Sources which show broad (FWHM $>
1000$~km s$^{-1}$) emission
lines (i.e. QSOs and type 1 Seyferts) are classed as broad-line AGN
(BLAGN). Sources which show only narrow emission lines are classed as narrow
emission line galaxies (NELGs). Sources for which no emission lines can be
discerned in the optical spectra are
classed simply as `galaxies'. Table \ref{tab:IDs} provides the current optical
identification statistics. 
The redshift distributions of the
sources are shown in Fig.\ref{fig:redshifts}. The BLAGN cover a wide range
of redshifts, but the NELGs and galaxies all have $ z < 1 $. 

For all those identified sources with $> 100$ EPIC counts we have constructed
and modeled the X-ray spectra (see Fig. \ref{fig:spectra}). To each spectrum we
have fitted an absorbed power law model. We have classified a source as
absorbed if either (i) the fit without intrinsic absorption was rejected at 
$>99\%$
confidence and was acceptable when absorption was included or (ii) the 
$\chi^{2}$ of the fit was significantly better with intrinsic absorption 
(at $ >99\%$ significance according to the F-test) than without.
The best fit photon indices are shown as a function of luminosity in Fig. 
\ref{fig:specresults}.
Some of
the spectra are certainly more complex than an absorbed power law, requiring
e.g. an additional soft 
component (as seen in Fig. \ref{fig:spectra}c); 
in these cases the photon index shown in
Fig. \ref{fig:specresults} is for the dominant power law component. 
To calculate the luminosities of the sources we have used the 2-5 keV flux 
measured with EPIC because this band requires only a small correction for
intrinsic absorption in most cases; this correction is shown as a horizontal
line for absorbed sources in Fig. \ref{fig:specresults}. 
All the luminosities were K-corrected using
the best fit spectral slopes, and we assumed 
$H_{0} = 50$~km~s$^{-1}$~Mpc$^{-1}$ and $q_{0}$=0.5. 

\section{Discussion}

As seen in Fig. \ref{fig:specresults}, the underlying spectral indices of the 
sources are
scattered around the mean value of $\Gamma \sim 2$ with a standard deviation of
$\sim 0.5$, and show no 
obvious dependence on luminosity, absorption, or optical classification. 
Furthermore, the distribution of spectral indices is remarkably similar to 
that 
found in brighter, soft X--ray selected samples (e.g. Mittaz et
al. 
\nocite{mittaz99} 1999). Thus the {\em XMM-Newton} 
spectroscopy is consistent with
a picture in which the underlying power law X-ray spectra of AGN do not 
depend on redshift or luminosity, and are the same in the absorbed and
unabsorbed sources. 

In the context of the `standard' picture of the X-ray populations, the majority
of the NELGs and galaxies, and certainly 
those with $L_{X} > 10^{42}$~erg~s$^{-1}$,
are expected to be intrinsically absorbed AGN.
As seen in Table \ref{tab:IDs} and
Fig. \ref{fig:specresults}, absorption is detected in $\sim 50\%$ of the  
galaxies and NELGs compared to only $\sim 10\%$ of the BLAGN; these latter
sources have $z > 1$ and column densities between $5\times 10^{21}$~cm$^{-2}$ and
$2\times 10^{22}$~cm$^{-2}$ and so are similar to the luminous, X-ray 
absorbed, BLAGN found in the hard-spectrum {\em ROSAT}
survey of Page, Mittaz \& Carrera \nocite{page01} (2001). 
Thus a significant fraction of the NELGs and galaxies are indeed absorbed AGN,
as reported for the Lockman Hole (Mainieri et al. \nocite{mainieri02} 2002).

However, a significant fraction ($\sim 50\%$) of the galaxies and NELGs do not
show strong evidence for X--ray absorption, but have power-law (i.e. AGN-like)
X--ray spectra and the majority of them 
have AGN X--ray luminosities.  The lack of broad
emission lines in these sources is not readily explained as an absorption
effect: assuming a Galactic dust to gas ratio, a column density of $5 \times
10^{21}$ cm$^{-2}$ is required to attenuate broad lines such as H$\beta$ by a
factor of 10. Such column densities are readily detectable for the (mostly
low-$z$) galaxies and NELGs. 

\begin{figure}
\resizebox{\hsize}{!}
{\includegraphics[]{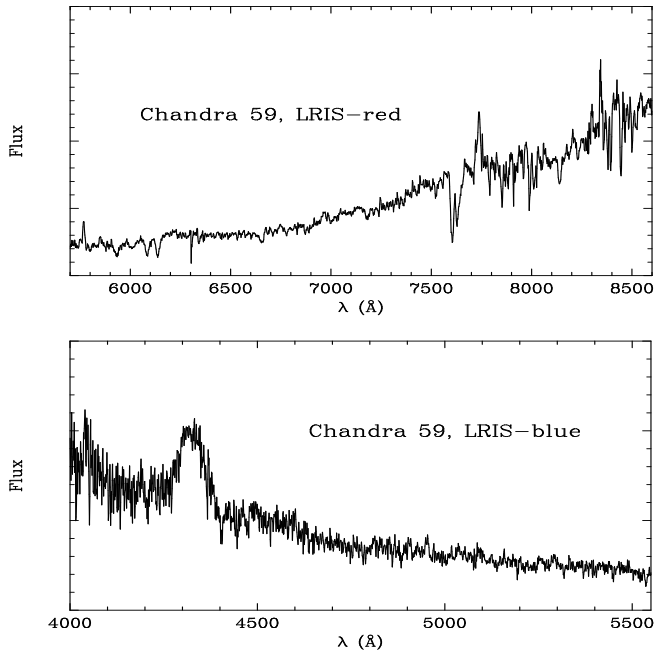}}
\caption{Keck LRIS red and blue arm spectra of Chandra~59, a BLAGN with 
z=0.546. In the red, the spectrum is dominated by the bright host galaxy but in
the blue (rest-frame UV) the AGN dominates and the broad Mg~II emission line 
is obvious}
\label{fig:chandra59}
\end{figure}

One feature that is noticeable in Fig. \ref{fig:specresults} is 
that the X--ray sources 
classified as NELGs and galaxies typically 
have
lower X--ray luminosities than the BLAGN. 
This is true for both the X--ray absorbed and unabsorbed
 NELGs and galaxies, even after corrections for absorption: 
{\it none} of the NELGs or galaxies have 
log~$L_{X} > 43.7$, but the {\it majority} of the BLAGN have 
log~$L_{X} > 43.7$. A likely explanation for the lack of broad 
emission lines
in the unabsorbed galaxy and NELG sources is therefore 
simply one of contrast: these
sources are weak (e.g. factor 10 underluminous) AGN 
in bright host galaxies. In fact even amongst the sources
which {\em are} classified as BLAGN, we can see this kind of phenomenon:
Fig. \ref{fig:chandra59} shows one such source, which is only recognisable as a
BLAGN in the rest frame UV where the host galaxy is weak. Thus the optical
classification of the lower luminosity AGN is likely to have a complex
dependancy on both redshift and the host-galaxy luminosity. However, the
presence of significant power law components in the X--ray spectra of NELGs 
and galaxies demonstrates that they do contain AGN; this has important
implications for the evolution of the AGN X--ray luminosity function (e.g. see
Page et al. \nocite{page97} 1997). Similar hypotheses of AGN/host-galaxy
contrast have also been proposed to
hide emission lines in Seyfert 2 galaxies (Lumsden \& Alexander
\nocite{lumsden01} 2001; Moran, Filippenko \& Chornock \nocite{moran02} 2002).
Another mechanism which can make broad
emission lines undectable is beamed continuum emission from BL Lacs; there are
at least two examples of this type of object in the 13hr field (see Gunn et
al. this proceedings).

If the lack of broad lines in X--ray unabsorbed AGN is a consequence of the
relative brightness of AGN and their host galaxies, the cosmological evolution
of AGN may offer an explanation as to why a large fraction of NELGs and
galaxies (and the equivalent sources in Rosati et
al. \nocite{rosati02} 2002 and Mainieri et al. \nocite{mainieri02} 2002) are
found predominantly at $z < 1$. The characteristic luminosity (i.e. the break
in the luminosity function) of AGN has declined dramatically (by a factor
$>10$) since $z=2$ at both X--ray (e.g. Page et al. \nocite{page97} 1997) and
optical wavelengths (e.g. Boyle et al. \nocite{boyle00} 2000). Unless the host
galaxies of AGN have declined in luminosity by a similar amount, the contrast
between AGN and host galaxy will typically be smaller today than it was at high
redshift.

\acknowledgements Based on observations obtained with {\em XMM-Newton}, an ESA
science mission with instruments and contributions directly funded by ESA
Member States and the USA (NASA).

\end{document}